\documentstyle[11pt]{article}
\begin{document}

\baselineskip=4.5mm

\vspace*{25mm}
\begin{center}
{\bf \Huge What do $\gamma$-ray bursts look like?}\footnote[1]{\em Invited 
talk, to appear in the Proceedings of the 1999 Pacific Rim Conference on 
Stellar Astrophysics.}
\end{center}
\vspace{5mm}
\begin{center}
{\large \bf T. Lu}

\vspace{2mm}
{\small Department of Astronomy, Nanjing University, Nanjing 210093, China}
\end{center}

\begin{abstract}
There have been great and rapid progresses in the field of $\gamma$-ray
bursts 
(denoted as GRBs) since BeppoSAX and other telescopes discovered their 
afterglows in 1997.
Here, we will first give a brief review on the observational facts of 
GRBs and direct understanding from these facts, which lead to the standard 
fireball model. The dynamical evolution of the fireball is discussed,
especially
a generic model is proposed to describe the whole dynamical 
evolution of GRB remnant from highly radiative to adiabatic, and from 
ultra-relativistic to non-relativistic phase. Then, Various deviations from
the 
standard model are discussed to give new 
information about GRBs and their environment. In order to relax the energy
crisis, 
the beaming effects and their possible observational evidences are also 
discussed in GRB's radiations. 
\end{abstract}
  

\section{Introduction}          

$\gamma$-ray burst (or shortly, GRB) is an astronomical phenomenon of short 
duration enhancement of $\gamma$-rays from cosmic space. It was discovered 
by R.W. Klebesadel et al. in 1967 and published in 1973. There have been more 
than 2000 GRBs discovered now. If there are appropriate satellites in the
sky, 
people could discover one or two GRBs everyday in average. However, they 
are still among the most mysterious astronomical 
objects even at present time (M\'{e}sz\'{a}ros 1999; Piran 1999a). The
difficulty encountered in this field lies in that GRBs occur at random,
people 
can not prepare to observe
in advance, they are too short in duration to be studied in detail, they were 
observed only in $\gamma$-ray band which has very low precision in
localization 
and could not be identified or associated with other objects of known
distances. 
Without the 
knowledge of distances, no serious studies could be made in astrophysics.

In early 1997, the Italian-Dutch Satellite, BeppoSAX, brought the great
break-through by rapid and accurate GRB localization and thus provided a 
small arcmin (or even smaller) error box. With so small an error box, 
it identified an X-ray counter-part (now known as X-ray afterglow) 
of GRB970228 even 8 hours after the $\gamma$-ray trigger (Costa et al. 1997). 
Several hours later, its optical afterglow was also observed 
(Groot et al. 1997; van Paradijs et al. 1997).
Since then, BeppoSAX has observed more than 20 GRBs, of which almost all 
exhibited X-ray afterglows. Based on the precise localization, many 
telescopes observed about a dozen optical afterglows and about ten radio 
afterglows. Up to now, people have observed host galaxies of more than 
ten GRBs with large red-shifts, showing them definitely at cosmological 
distances. These great discoveries lead to rapid developments. A lot of 
questions have now been clarified. However, compared with GRB itself,
afterglow 
appears to be simpler and has been known much better. In contrast, the GRB 
itself, especially its energy source and origin, still keeps to be mysterious.

The main observational facts of GRBs are as follows (Piran 1999a):

{\bf Temporal properties:}
 The GRB duration ($T$) is very short, usually only a 
few seconds or tens of seconds, or occasionally as long as a few tens of
minutes
or as short as a few milli-seconds. Their time profiles are diverse, some
may be 
simply shaped, others complicated, multi-peaked. There seems to be a roughly 
bimodal distribution of long bursts of $T \geq 2$ s and short bursts 
of $T \leq 2$ s. The time scales of variability 
($\delta T$), especially their rising time, may be as short as only
milli-seconds or 
even sub-milli-seconds. Typically, $\delta T \sim 10^{-2} T$.

{\bf Spectral properties:}
The photon energy radiated in GRB is typically in the range of tens 
keV to a few MeV. However, high energy tail up to GeV or even higher than
10 GeV 
does exist. The spectra are definitely non-thermal and can usually be fitted 
by power law $N(E)dE \propto E^{-\alpha} dE$ (or break power law) with 
index $\alpha$ within about $1.8$ to $2.0$. The $\gamma$-ray fluences 
are typically in the range of $(0.1 - 10) \times 10^{-6}$ ergs/cm$^{2}$.

{\bf Spatial distribution:} After the launch of the CGRO (Compton Gamma-Ray 
Observatory) Satellite in 1991, the Burst and Transient Source Experiment
(BATSE)  
showed clearly that the spatial 
distribution of GRB sources is highly isotropic with almost zero dipole and
quadrupole components (Meegan et al. 1992). This distribution favors GRBs
at cosmological 
distances at least statistically. However, GRBs at extended dark halo of our 
Galaxy could also explain this feature. This led to a great debate between 
Galactic origin and cosmological origin. 

{\bf Afterglows:} Afterglows are counterparts of GRBs at wave bands other
than 
$\gamma$-rays, may be in X-ray, optical, or even radio bands. They are
variable, 
typically decaying according to power laws: $F_{\nu} \propto t^{-\alpha}$ 
($\nu=$ X, optical, ......) with $\alpha = 1.1 - 1.6$ for X-ray, $\alpha 
= 1.1 - 2.1$ for optical band. X-ray afterglows can last days or even weeks; 
optical afterglows and radio afterglows months. The most important discovery 
is that many 
afterglows show their host galaxies being definitely at cosmological
distances 
(with large red-shifts up to $Z=3.4$ or even 5). Thus, the debate is 
settled down, GRBs are at cosmological distances, they should be the most 
energetic events ever known since the Big Bang.

\section{The Standard Fireball Shock Model}

{\bf Stellar level event:} The variability time scale is usually very
short. Let 
$\delta T \sim {\rm ms}$, then, the space scale of the initial source, 
$R_{\rm i} <c\delta T \sim 3 \times 10^{2} {\rm km}$. Hence, even for black
hole, 
considering $R = 2{\rm G}M/c^{2}$, we have 
$M \leq \frac{{\rm c}^{3}\delta T}{2{\rm G}} \sim 10^{2}$~M$_{\odot}$. If the 
burster is not a black hole, its mass should be much smaller. Thus, we can
conclude 
that the GRB should be 
{\bf a stellar phenomenon} and the burster should be {\bf a compact 
stellar object} which may be related with neutron star (or strange star) or 
stellar black hole. 

{\bf Fireball:} From the measured fluence $F$ and the measured distance $D$, 
if emission is 
isotropic, we can calculate the total radiated energy to be $E_{0} = F(4\pi 
D^{2})\approx 10^{51}(F/(10^{-6}{\rm ergs/cm}^{2}))(D/(3{\rm Gpc}))^{2}$.
Thus, 
very large energy ($10^{51}$ ergs) is initially contained in a small volume
of 
$(4/3)\pi R_{i}^{3} \sim 1 \times 10^{23}$ cm$^{3}$. This should be
inevitably 
a fireball, of which the optical depth for 
$\gamma \gamma \longrightarrow {\rm e}^{+} {\rm e}^{-}$,
$\tau_{\gamma\gamma}$, 
is very large. Consider a typical burst with an observed fluence 
$F$, at a distance $D$, with a temporal variability time scale 
$\delta T$, its average optical depth can be written as: 
\begin{equation}
\tau_{\gamma\gamma} = \frac{f_{\rm p} \sigma_{\rm T} F D^{2}}{R_{\rm i}^{2}
m_{\rm e} c^2}
\sim 10^{17} f_{\rm p}
\bigg(\frac{F}{10^{-6} {\rm ergs/cm^2}}\bigg)
\bigg(\frac{D}{3~{\rm Gpc}}\bigg)^2
\bigg(\frac{\delta T}{1~{\rm ms}}\bigg)^{-2},  
\end{equation}
where $f_{\rm p}$ denotes the fraction of photon pairs 
satisfying $\sqrt{E_1 E_2} > $m$_{\rm e}$ c$^2$. 

For so large an optical depth, there seem to appear two serious difficulties. 
First, the radiation in an optically thick case should be thermal, while 
the observed radiation is definitely non-thermal. Second, high energy 
photons should be easily converted into ${\rm e}^{+} {\rm e}^{-}$ pairs, 
while the observed high energy tail indicates that 
this convertion has not happened. However, it is very interesting to note 
that just such a large optical depth paves the way to solve both of them.

{\bf Compactness problem:} In fact, the luminosity of the thermal 
radiation, according to the 
Stefan-Boltzmann law, should be proportional to the surface of the 
fireball which is initially so small that the thermal radiation can not 
be observed. However, just due to the large optical depth, the radiation 
pressure should be very high and could accelerate the fireball expansion 
to become ultra-relativistic with a large Lorentz factor $\gamma$. After 
expanding to a large enough distance, it may be getting optically thin.
At this time, the non-thermal $\gamma$-ray bursts can be observed. Does 
such a large distance contradict the compactness relation 
$R_{\rm i} \le {\rm c} \delta T$ compared with the 
milli-second variabilities? To answer this question, let us first 
note that this relation holds only for non-relativistic (rest) 
object with $R_{\rm i}$ denoting its space scale. For an
ultra-relativistically 
expanding fireball, the compactness relation should be relaxed to 
\begin{equation}
R_{\rm e} \le \gamma^{2} {\rm c} \delta T,
\end{equation}
here $R_{\rm e}$ is the space scale of the expanding fireball with Lorentz 
factor $\gamma$. Considering two photons we observed at two different times 
apart by $\delta T$, as the emitting region is moving towards the observer 
with a Lorentz factor $\gamma \gg 1$, the second photon
should be emitted at a far nearer place than the first one. This gives 
effectively short time variabilities and leads to the additional factor 
$\gamma^{2}$ appearing in the above compactness relation.

The factor $f_{\rm p}$ in the optical depth $\tau_{\gamma\gamma}$ also 
sensitively depends on the ultra-relativistic expansion of the fireball. 
As for this case, the observed photons are blue-shifted, in the comoving 
frame, their energy 
should be lower by a factor of $\gamma$, and fewer photons will have 
sufficient energy to produce pairs. This gives a factor depending 
on spectral index $\alpha$, namely a factor of $\gamma^{2\alpha}$ in 
$\tau_{\gamma\gamma}$. 

{\bf Ultra-relativistic expansion:} Therefore, the optical 
depth $\tau_{\gamma\gamma}$ will decrease by a 
factor of $\gamma^{4+2\alpha}$ 
for the ultra-relativistically expanding fireball (Goodman 1986; 
Paczy\'{n}ski 1986; Piran 1999a; Krolik \& Pier 1991):
\begin{equation}
\tau_{\gamma\gamma} =
{f_{\rm p} \over \gamma ^{2 \alpha}} {{\sigma_{\rm T} F D^2} 
  \over {R_{\rm e}^2 {\rm m}_{\rm e} {\rm c}^2}}
\approx {10^{17} \over \gamma ^{(4+2\alpha)}}
f_{\rm p}
\bigg({ F \over 10^{-6} {\rm ergs/cm^2}} \bigg)
\bigg({ D \over 3~{\rm Gpc}}\bigg)^2
\bigg({ \delta T \over 1~{\rm ms}} \bigg)^{-2}.
\end{equation}
Note, the spectral index $\alpha$ is approximately 2, we will have 
$\tau_{\gamma\gamma} < 1$ for $\gamma > 10^{17/(4+2\alpha)} \sim 10^{2}$. 
Thus, in order for the fireball to become optically thin, as required by 
the observed non-thermal spectra of $\gamma$-ray bursts, its expanding 
speed should be ultra-relativistic with Lorentz factor 
\[\gamma > \sim 10^{2}.\]
This is a very important character for GRBs, which 
limits the baryonic mass contained in the fireball seriously. If the initial
energy is $E_{0}$, then the baryonic mass $M$ should be less than 
\begin{equation}
E_{0}/({\rm c}^{2}\gamma) \le 10^{-5}{\rm M}_{\odot} (E_{0} / (2\times
10^{51} {\rm ergs})),
\end{equation}
otherwise, the initial energy can not be converted to the kinetic energy of 
the bulk motion of baryons with such a high Lorentz factor. Most models
related 
with neutron stars contain baryonic mass much higher than this limit. This is 
the famous problem named as ``baryon contamination''.

It is worthwhile to note that this very condition $\gamma > \sim 10^{2}$ can
also explain the existence of the high energy tail in the GRB spectra, as the 
observed high energy
photons should be only low energy photons in the frame of emitting region,
they
are not energetic enough to be converted into ${\rm e}^{+}{\rm e}^{-}$ pairs.

{\bf Internal-external shock:} What is the radiation mechanism in the
fireball 
model? The fireball 
expansion has successfully made a conversion of the initial internal energy 
into the bulk kinetic energy of the expanding ejecta. However, this is the 
kinetic energy of the associated protons, not the photons. We should have 
another mechanism to produce radiation, otherwise, even after the fireball 
becoming optically thin, the $\gamma$-ray bursts can not be observed. 
Fortunately, the shocks described below can do such a job. 

The fireball can be regarded as roughly homogeneous in its local rest frame, 
but due to the Lorentz contraction, it looks like a shell (ejecta) with 
width of the initial size of the fireball. As the shell collides with 
inter-stellar medium (ISM), shocks will be 
produced (Rees \& M\'{e}sz\'{a}ros 1992; Katz, J., 1994; 
Sari \& Piran 1995; Mitra 1998). This is usually 
called as external shocks. Relativistic electrons that have been 
accelerated in the relativistic shocks will usually emit synchrotron 
radiation. As the amount of swept-up interstellar matter getting larger 
and larger, the shell will be decelerated and radiation of longer wave 
length will be emitted. Thus, an external shock can produce only smoothly 
varying time-dependent emission, not the spiky multi-peaked structure 
found in many GRBs. If the central energy source is not completely 
impulsive, but works intermittently, it can produce many shells (or 
many fireballs) with different Lorentz factors. Late but faster shells 
can catch up and collide with early slower ones, and then, shocks 
(internal shocks) thus produced will lead to the observed bursting 
$\gamma$-ray emission (Rees \& M\'{e}sz\'{a}ros 1994; 
Paczy\'{n}ski \& Xu 1994). This 
is the so called external-internal shock model, 
internal shocks give rise to $\gamma$-ray bursts and external shocks to 
afterglows. The internal shocks can only convert a part of their energies 
to the $\gamma$-ray bursts, other part remains later to interact with the 
interstellar medium and lead to afterglows. Typically, the GRB is produced 
at a large distance of about 10$^{13}$ cm to the center, such a large 
distance is allowed according to the relaxed compactness relation 
$R_{\rm e} \le \gamma^{2} {\rm c} \delta T$, while its afterglows are
produced at 
about 10$^{16}$ cm or even much farther. This internal-external shock 
scenario, under the simplified assumptions of uniform 
environment with typical ISM number density of $n \sim 1  {\rm cm^{-3}}$, 
isotropic emission of synchrotron radiation and only impulsive energy 
injection, is known as the standard model.  

{\bf Spectra of afterglows:} The instantaneous spectra of afterglows,
according 
to this model, can be 
written as $F_{\nu} \propto \nu^{\beta}$, with different $\beta$ for 
different range of frequency $\nu$ (Sari et al. 1998; Piran 1999b). Let 
$\nu_{sa}$ be the self absorption frequency, for which the optical depth 
$\tau(\nu_{sa})=1$. For $\nu < \nu_{sa}$, we have the Wien's law: 
$\beta=2$. For $\nu_{sa} < \nu < \min~(\nu_{\rm m},\nu_{\rm c})$, we 
can use the low energy synchrotron tail, $\beta=-1/3$. Here 
$\nu_{\rm m}$ is the synchrotron frequency of an electron with 
characteristic energy, $\nu_{\rm c}$ is the cooling frequency, namely 
the synchrotron frequency of an electron that cools during the local 
hydrodynamic time scale. For frequency within $\nu_{\rm m}$ and 
$\nu_{\rm c}$, we have $\beta = -1/2$ for fast cooling 
($\nu_{\rm c} < \nu_{\rm m}$) and $\beta=-(p-1)/2$ for slow cooling 
($\nu_{\rm m} < \nu_{\rm c}$). For $\nu > \max(\nu_{\rm m},\nu_{\rm c})$, 
we have $-p/2$. Here, $p$ is the spectral index of the emitting 
electrons: $N(E)\propto E^{-p}$.

\section{Dynamical Evolution of the Fireball}

During the $\gamma$-ray bursting phase and the early stage of afterglows,
the fireball expansion is initially ultra-relativistic and highly 
radiative, but finally it would be getting into non-relativistic and
adiabatic, 
a unified dynamical evolution should match all these phases. In fact, the 
initial ultra-relativistic phase has been well described by some simple
scaling 
laws (M\'{e}sz\'{a}ros \& Rees 1997a; Vietri 1997; Waxman 1997; Wijers et
al. 1997), 
while the final 
non-relativistic and adiabatic phase should obey the Sedov (1969) rule, 
which has well been studied in Newtonian approximation. The key equation
(Blandford \& McKee 1976; Chiang \& Dermer 1999) is
\begin{equation}
\frac{{\rm d}\gamma}{{\rm d}m} = -\frac{\gamma^{2}-1}{M},
\end{equation}
here $m$ denotes the rest mass of the swept-up medium, $\gamma$ the bulk
Lorentz 
factor, and $M$ the total mass in the co-moving frame including internal
energy $U$. 
This equation was originally derived under the ultra-relativistic
condition. The 
widely accepted results derived under this equation are correct for 
ultra-relativistic expansion. Accidentally, these results are also suitable
for the 
non-relativistic and radiative case. However, for the non-relativistic and 
adiabatic case, they will lead to wrong result ``$v \propto R^{-3}$'' ($v$
is the 
velocity), while the correct Sedov result should be ``$v \propto R^{-3/2}$'', 
as first pointed out by Huang, Dai and Lu (1999a,b).

It has been proved (Huang, Dai \& Lu 1999a,b) that in the general case, 
the above equation should be replaced by 
\begin{equation}
\frac{{\rm d}\gamma}{{\rm d}m} = -\frac{\gamma^{2}-1}{M_{ej} + \epsilon m 
+ 2(1-\epsilon)\gamma m},
\end{equation}
here $M_{ej}$ is the mass ejected from GRB central engine, $\epsilon$ is the 
radiated fraction of the shock generated thermal energy in the co-moving 
frame. The above equation will lead to correct results for all cases 
including the Sedov limit. This generic model is suitable for both 
ultra-relativistic and non-relativistic, and both radiative and adiabatic 
fireballs. As proved by Huang et al. (1998a,b), Wei \& Lu (1998a) and 
Dai et al. (1999a), 
only several days after the burst, a fireball will usually become 
non-relativistic and adiabatic, while the afterglows can last some months, 
the above generic model is really useful and important.

\section{Comparison and Association of GRB with SN}

Supernova was known as the most energetic phenomenon at the stellar level. 
SN explosion is the final violent event in the stellar evolution. 
Dynamically, it can also be described as a fireball, which however expands 
non-relativistically. After the SN explosion, there is usually a remnant
which 
can shine for more than thousands of years and be well described
dynamically by 
Sedov model (Sedov 1969).

GRB is also a phenomenon at the stellar level. However, it is much more 
energetic and much more violent than SN explosion! It has been proved to 
be described 
as a fireball, which expands ultra-relativistically. The GRB may also leave
a remnant which shines for months now known as afterglow. 

Their comparison is given in Table I:
\begin{center}
\begin{tabular}{|l|l|l|}
\multicolumn{3}{c}{Table I} \\ \hline
      &  GRBs    &    SNs    \\   \hline
  {\bf Burst}  &  {\bf Bursting $\gamma$-rays}  &   {\bf SN explosion}  \\
\hline   
  Energy up to     &  10$^{54}$ ergs &   $10^{51}$ ergs   \\
  Time Scale &  10 sec &   Months   \\
  Profile  &  irregular   &   smooth \\
  Wave Band  &  $\gamma$-ray  &  Optical \\  \hline
{\bf Relic}  &  {\bf Afterglow}  &  {\bf Remnant} \\ \hline
Time Scale  &  Months  &  10$^{3}$ Years \\
Wave Band  &  Multi-band  &  Multi-band  \\ \hline
{\bf Understanding}  &  &  \\  \hline
Fireball Expansion  &  Ultra-relativistic  &  Non-relativistic  \\
Mechanism  &  ???  &  Stellar Core Collapse  \\
Key Process  &  ???   &   Neutrino process  \\  \hline
\end{tabular}
\end{center}

In April 1998, a SN 1998bw was found to be in the 8' error circle of the X-ray
afterglow of GRB 980425 (Galama et al. 1998; Kulkarni et al. 1998). However, 
its host galaxy is at a red-shift z=0.0085 (Tinney et al. 1998), indicating a 
distance of 38 Mpc (for $H_{0} = 65$ km s$^{-1}$ Mpc$^{-1}$), which leads the 
energy of the GRB to be too low, only about 
$5 \times 10^{47}$ ergs, 4 orders of magnitude lower than normal GRB. 

Later, in the light curves of GRB 980326 (Bloom et al. 1999; 
Castro-Tirado \& Gorosabel 1999b) and GRB 970228 (Reichart 1999; 
Galama et al. 1999b), some evidence related with SN was found. This is 
a very important question worth while to study further (see e.g. Wheeler 
1999). These two violent phenomena, GRB and SN, might be closely related. 
They might be just two steps of one single event (Woosley et al. 1999; 
Cheng \& Dai 1999; Wang et al. 1999b; Dai 1999d). It is interesting to 
note that the first step might provide a low baryon environment for 
the second step to produce GRB. Such a kind of models can give a way 
to avoid the baryon contamination.

\section{Inner Engine and Energetics}

There have been a lot of models proposed to explain the central engine of
GRBs 
(see e.g. Castro-Tirado 1999a; Piran 1999a; Cheng \& Dai 1996; 
Dai \& Lu 1998b). 
All these objects are related with compact stars such as neutron star (NS), 
strange star (SS), black hole (BH) etc. For example, binary mergers (NS-NS, 
NS-BH, ...), massive star collapsing, phase transitions (NS to SS) and others 
have been proposed. To build a successful model for central engine, the most 
difficult task is to solve the baryon contamination. There seem to be three 
kinds of ways: 1) based on BH, which can swallow baryons; 2) based on SS, of 
which baryons are only contained in its crust with mass less than 
$10^{-5}$ M$_{\odot}$; 3) based on the two-step process pointed out in 
above section.

A system of a central BH with a debris torus rotating round it may form after 
compact star merging or massive star collapsing. In this system, two kinds of 
energies can be used: the rotational energy of the BH and the gravitational 
energy of the torus. The rotational energy of the BH can be extracted via the 
B-Z (Blandford \& Znajek 1977) mechanism (M\'{e}sz\'{a}ros \& Rees 1997b;
Paczy\'{n}ski 1998). For 
a maximally rotating BH, its rotational energy can be extracted up to 29\% of 
the BH rest mass, while the gravitational binding energy of the torus can be 
extracted up to 42\% of the torus rest mass. Lee, Wijers and Brown (1999) 
recently studied the possibility to use these mechanisms in producing GRB. 

The phase transition from neutron star to strange star can release huge
energy 
to account for GRB. As an estimate, we can reasonably assume that about 20-30 
MeV is released per baryon during the phase transition. Total energy released 
this way can be up to about $(4-6) \times 10^{52}$ ergs. Strange star is the 
stellar object in the quark level. Whether it exists or not is a fundamental 
physical/astrophysical problem. Its main part is a quark core with large 
strangeness (known as strange core). There could be a thin crust with mass of 
only about $\sim (10^{-6} - 10^{-5})$M$_{\odot}$ (Alcock et al. 1986; Huang
\& Lu 
1997a,b; Lu 1997; Cheng et al. 1998), all baryons are contained in the crust. 
It is interesting to note that this baryonic mass is low enough to
avoid the baryon contamination. Klu\'{z}niak and Ruderman (1998) proposed 
differentially rotating neutron stars as an origin of GRBs. Dai and Lu
(1998b) 
used this mechanism to the case of differentially rotating strange stars and 
proposed a possible model for GRB without baryon contamination.

\section{New Information Implied by the Deviations from the Standard Model}

The standard model described above is rather successful in that its physical 
picture is very clear, it gives results very simple, and observations on GRB
afterglows support it at least qualitatively but generally. However, various 
quantitative deviations have been found. They indicate that the 
simplifications made in the standard model should be improved. These 
deviations may reveal important new information, such as non-uniform 
environment, additional energy injection, beaming effects of radiation 
and others.

{\bf Wind environment effects:} Dai and Lu (1998c) analysed the afterglows 
of GRB970616 and others. They 
studied the general case of $n \propto R^{-k}$ for the non-uniform
environment 
density (here $n$ is the number density of the environment medium) and 
found that the X-ray afterglow of GRB970616 can well be fitted by $k = 2$, 
and pointed out that this non-uniformity may be due to the existence of 
a stellar wind. After the detailed studies by Chevalier \& Li (1999a,b), the 
stellar wind model has become widely interested. People are aware that it may 
contain many implications about the pregenitors of GRBs and provide
strong support to their massive star origin.

{\bf Additional energy injection:} The optical afterglows of GRB 970228 and 
GRB 970508 had some complexities, 
showing down-up-down variation in their light curves. These features can be 
explained by long time scale energy injection from their central engines 
(Dai \& Lu 1998a,b; Rees \& M\'{e}sz\'{a}ros 1998; Panaitescu et al. 1998). 
For example, a milli-second pulsar with 
super-strong magnetic field can be produced at birth of GRB. As the 
fireball expands, the central pulsar can continuously 
supply energy through magnetic dipole radiation. Initially, the energy 
supply is small enough, the afterglow shows declining. As it becomes 
important, the afterglow shows rising. However, the magnetic dipole 
radiation should itself attenuate later. Thus, the down-up-down shape 
would appear naturally. Dai and Lu (1999c) analysed GRB 980519, 990510 
and 980326, and obtained the results agreeing well with observations.

{\bf Additional radiations:} Though the synchrotron radiation is 
usually thought to be the main radiation mechanism, 
however, under some circumstances, the inverse Compton scattering may play an 
important role in the emission spectrum, and this may influence the temporal 
properties of GRB afterglows (Wei \& Lu 1998a,b).

Later data in the afterglows of GRB 970228 (Reichart 1999; 
Galama et al. 1999b) and 980326 (Bloom et al. 1999; Castro-Tirado \& 
Gorosabel 1999b) may show the deviations as additional contributions
from supernovae.

{\bf Beaming effects:} 
GRB 990123 has been found very strong in its 
$\gamma$-ray emission, and the red 
shift of its host galaxy is very large (z=1.6) (Kulkarni et al. 1999a; 
Galama et al. 1999a; Akerlof et al. 1999; Castro-Tirado, et al. 1999a; 
Hjorth, et al. 1999; Andersen, et al. 1999). If its radiation is isotropic, 
the radiation energy only in $\gamma$-rays is already as high as  
$E_{\gamma} \sim 3.4 \times 10^{54}$ ergs, closely equals two solar rest 
energy ($E_{\gamma} \approx 2{\rm M}_{\odot}$c$^{2}$)! As the typical mass
of the 
stellar object is in the order of $\sim 1 {\rm M}_{\odot}$, while the
radiation 
efficiency for the total energy converting into the $\gamma$-ray emission 
is usually very low, such a high emission energy is very difficult to 
understand (Wang, Dai \& Lu 1999a).

A natural way to relax the energy crisis is to assume that the radiation of
GRB 
is beaming, rather than isotropic. Denote the jet angle as $\Omega$, 
then the radiation energy $E$ will be reduced to $E\Omega/4\pi$. At the
same time, 
the estimated burst rate should increase by a factor of $4\pi/\Omega$. Are 
there any observational evidences for the jet in GRB and its afterglow? 
Pugliese et al. (1999), Rhoads (1997, 1999), Sari et al. (1999) and 
Wei \& Lu (1999a,b) have discussed this question. 
Kulkarni et al. (1999b) observed that two days after the burst, the
decaying was 
getting more steepening, appearing as a break in the light curve of GRB
990123, 
and they regarded this as the evidence for jet. Recently, Huang et al.
(1999c) 
calculated the influences of various parameters on the jetted emission of
GRB, 
showed that a break in the light curve may appear in the case of narrow jet
and 
for small electron energy fraction and small magnetic energy fraction. 

{\bf Dense environment effects:} However, Dai and Lu (1999b) 
pointed out that a shock undergoing the transition from a relativistic
phase to 
a non-relativistic phase may also show a break in the light curve, if there
are 
dense media and/or clouds in the way, this break may happen earlier. This
model 
could also give an explanation for the observed steepening. Recently, 
Wang et al. (1999c) proved that the dense environment model can also 
explain well the radio afterglow of GRB 980519 (Frail et al. 1999). Thus, we 
should study the break appearing in the light curve further to take both 
beaming and dense environment effects into account. 

\section{Conclusion}

The standard internal-external shock model, which is built under many 
simplifications, has been proved to be well fitted by 
observations qualitatively but generally. Based on the success of this 
model, it should be very important to study the deviations from the 
standard model, which indicate that the simplifications should be 
relaxed in some aspects. Hence, the deviations contain 
important new information and have been a fruitful research area. 

In contrast to the rapid progress in understanding the nature of 
afterglows, GRB itself has not yet been clear. However, this is a very 
important problem. The solution about the origin of GRB is 
related with most fundamental questions in physics and astrophysics, such as 
black holes, stellar objects from quark level, physics and properties of the 
farthest stellar phenomena and others. It may also give new and important 
information for cosmology. Within five or ten years, there should still be
a lot 
of further surprising achievements about these most violent events. 

\vspace{4mm}

\noindent
{\bf Acknowledgments:} This work is supported by the National Natural 
Science Foundation of China.

\vspace{6mm}
\begin{center}
{\bf \large References:}
\end{center}
\baselineskip=0mm
\begin{enumerate}

\item{} Akerlof, C., et al., 1999, Nature, 398, 400.
\item{} Alcock, C., Farhi, E., Olinto, A., 1986, ApJ 310, 261.
\item{} Andersen, M.I., et al., 1999, Science, 283, 2075.
\item{} Blandford, R.D., MaKee, C.F., 1976, Phys. Fluids, 19, 1130.
\item{} Blandford, R.D., Znajek, R.L., 1977, MNRAS, 179: 433.
\item{} Bloom, J.S. et al., 1999, Nature, 401, 453. 
\item{} Castro-Tirado, A.J., et al., 1999a, Science, 283, 2069.
\item{} Castro-Tirado, A., Gorosabel, J., 1999b, astro-ph/9906031.
\item{} Cheng, K.S., Dai, Z.G., 1996, Phys. Rev. Lett., 77, 1210.
\item{} Cheng, K.S., Dai, Z.G., Lu, T., 1998, Int. J. Mod. Phys. D, 7, 139.
\item{} Cheng, K.S., Dai, Z.G., 1999, astro-ph/9908248.
\item{} Chevalier, R.A., Li, Z.-Y., 1999a, ApJ, 520, L29.
\item{} Chevalier, R.A., Li, Z.-Y., 1999b, astro-ph/9908272.
\item{} Chiang, J., Dermer, C.D., 1999, ApJ, 512, 699.
\item{} Costa, E., et al., 1997, Nature, 387, 783.
\item{} Dai, Z.G., Lu, T., 1998a, A\&A, 333, L87.
\item{} Dai, Z.G., Lu, T., 1998b, Phys. Rev. Lett., 81, 4301.
\item{} Dai, Z.G., Lu, T., 1998c, MNRAS, 298, 87.
\item{} Dai, Z.G., Huang, Y.F., Lu, T., 1999a, ApJ, 520, 634.
\item{} Dai, Z.G., Lu, T., 1999b, ApJ, 519, L155.
\item{} Dai, Z.G., Lu, T., 1999c, astro-ph/9906109.
\item{} Dai, Z.G., 1999d, in this Proceedings.
\item{} Frail, D.A., et al., 1999, astro-ph/9910060.
\item{} Galama, T.J., et al., 1998, Nature, 395, 670.
\item{} Galama, T.J., et al., 1999a, Nature, 398, 394.
\item{} Galama, T.J., et al., 1999b, ApJ, astro-ph/9907264.
\item{} Goodman, J., 1986, ApJ, 308, L47.
\item{} Groot, P.J. et al., 1997, IAUC, No.6584.
\item{} Hjorth, J., et al., 1999, Science, 283, 2073.
\item{} Huang, Y.F., Lu, T., 1997a, Chin. Phys. Lett., 14, 314.
\item{} Huang, Y.F., Lu, T., 1997b, A\&A, 325, 189-194.
\item{} Huang, Y.F., Dai, Z.G., Lu, T., 1998a, A\&A, 336, L69.
\item{} Huang, Y.F., Dai, Z.G., Wei, D.M., Lu, T., 1998b, MNRAS, 298, 459.
\item{} Huang, Y.F., Dai, Z.G., Lu, T., 1999a, MNRAS, 309, 513.
\item{} Huang, Y.F., Dai, Z.G., Lu, T., 1999b, Chin. Phys. Lett., 16, 775, 
astro-ph/9906404.
\item{} Huang, Y.F., Gou, L.J., Dai, Z.G., Lu, T., 1999c, astro-ph/9910493.
\item{} Katz, J., 1994, ApJ, 422, 248.
\item{} Klebesadel, R.W., Strong, I.B., Olson, R.A., 1973, ApJ, 182, L85.
\item{} Klu\'{z}niak, W., Ruderman, M., 1998, ApJ, 505, L113.
\item{} Krolik, J.H., Pier, E.A., 1991, ApJ, 373, 277.
\item{} Kulkarni, S. R., et al., 1998, Nature, 395, 663.
\item{} Kulkarni, S. R., et al., 1999a, Nature, 398, 389.
\item{} Kulkarni, S. R., et al., 1999b, astro-ph/9903441.
\item{} Lee, H.K., Wijers, R.A.M.J., Brown, G.E., 1999, astro-ph/9906213.
\item{} Lu, T., 1997, Pacific Rim Conference on Stellar Astrophysics,
A.S.P. Conf.
       Ser. Vol. 138, eds. K.L. Chan, K.S. Cheng, H.P. Singh, 1998, 215.
\item{} Meegan, C. A., et al., 1992, Nature, 355, 143.
\item{} M\'{e}sz\'{a}ros, P., Rees, M.J., 1992, MNRAS, 257, 29.
\item{} M\'{e}sz\'{a}ros, P., Rees, M.J., 1997a, ApJ, 476, 232.     %
\item{} M\'{e}sz\'{a}ros, P., Rees, M.J., 1997b, ApJ, 482, L29.
\item{} M\'{e}sz\'{a}ros, P., 1999, astro-ph/9904038.
\item{} Mitra, A., 1998, ApJ, 492, 677.
\item{} Paczy\'{n}ski, B., 1986, ApJ, 308, L51.
\item{} Paczy\'{n}ski, B., Xu, G., 1994, ApJ, 427, 708.
\item{} Paczy\'{n}ski, B., 1998, ApJ, 494, L45.
\item{} Panaitescu, A., M\'{e}sz\'{a}ros, P., Rees, M.J., 1998, ApJ, 503, 314.
\item{} Piran, T., 1999a, Phys. Rep., 314, 575.
\item{} Piran, T., 1999b, astro-ph/9907392.
\item{} Pugliese, G., Falcke, H., Biermann, P.L., 1999, astro-ph/9903036.
\item{} Reichart, D.E., 1999, ApJ, 521, L111.
\item{} Rees, M.J., M\'{e}sz\'{a}ros, P., 1992, MNRAS, 258, 41.
\item{} Rees, M.J., M\'{e}sz\'{a}ros, P., 1994, ApJ, 430, L93.
\item{} Rees, M.J., M\'{e}sz\'{a}ros, P., 1998, ApJ, 496, L1.
\item{} Rhoads, J., 1997, ApJ, 487, L1.
\item{} Rhoads, J., 1999, ApJ, 525, 737, (astro-ph/9903399).
\item{} Sari, R., Piran, T., 1995, ApJ, 455, L143.
\item{} Sari, R., Piran, T., Narayan, R., 1998, ApJ, 497, L17.
\item{} Sari, R., Piran, T., Halpern, J.P., 1999, ApJ, 519, L17.
\item{} Sedov, L., 1969, Similarity and Dimensional Methods in Mechanics 
       (Academic, New York), Chap.IV.
\item{} Tinney, C. et al., 1998, IAU Circ. 6896.
\item{} Van Paradijs, J., et al., 1997, Nature, 386, 686.
\item{} Vietri, M., 1997, ApJ, 488, L105.
\item{} Wang, X.Y., Dai, Z.G., Lu, T., 1999a, astro-ph/9906062.
\item{} Wang, X.Y., Dai, Z.G., Lu, T., Wei, D.M., Huang, Y.F., 1999b, 
       astro-ph/9910029.
\item{} Wang, X.Y., Dai, Z.G., Lu, T., 1999c, astro-ph/9912492.
\item{} Waxman, E., 1997, ApJ, 485, L5.
\item{} Wei, D.M., Lu, T., 1998a, ApJ, 499, 754.
\item{} Wei, D.M., Lu, T., 1998b, ApJ, 505, 252.
\item{} Wei, D.M., Lu, T., 1999a, astro-ph/9908273.
\item{} Wei, D.M., Lu, T., 1999b, astro-ph/9912063.
\item{} Wheeler, J.Craig, 1999, astro-ph/9909096.
\item{} Wijers, R.A.M.J., Rees, M.J., M\'{e}sz\'{a}ros, P., 1997, MNRAS,
288, L51.
\item{} Woosley, S.E., Macfadyen, A.I., Heger, A., 1999, astro-ph/9909034.

\end{enumerate}
\end{document}